\documentclass[aps,pre,twocolumn,groupedaddress,showpacs]{revtex4}  
\usepackage{graphicx}   
\usepackage{bm}   
\renewcommand{\v}[1]{$\mathbf{#1}$}
\newcommand{\taubf}{$\tau_{bf}$}
\newcommand{\taual}{$\tau_{al}$}
\newcommand{\rb}{$C_{bf}$}
\newcommand{\rbcr}{$C^{cr}_{bf}$}
\newcommand{\prcell}{programmed cell death}
\newcommand{\patho}{pathological process}

\begin{document}

\title{Simulation of Spread and Control of Lesions in Brain}
\author{Krishna Mohan, T. R.} 
\affiliation{CSIR Centre for Mathematical Modelling and Computer Simulation (C-MMACS)\\Bangalore 560017, India}
\email{kmohan@cmmacs.ernet.in}
\homepage{http://www.cmmacs.ernet.in/~kmohan}
\date{\today}

\begin{abstract}
A simulation model for the spread and control of lesions in the brain is constructed using a planar network (graph) representation for the Central Nervous System (CNS). The model is inspired by the lesion structures observed in the case of Multiple Sclerosis (MS), a chronic disease of the CNS. The initial lesion site is at the center of a unit square and spreads outwards based on the success rate in damaging edges (axons) of the network. The damaged edges send out alarm signals which, at appropriate intensity levels, generate programmed cell death. Depending on the extent and timing of the programmed cell death, the lesion may get controlled or aggravated akin to the control of wild fires by burning of peripheral vegetation. The parameter phase space of the model shows smooth transition from uncontrolled situation to controlled situation. The simulations show that the model is capable of generating a wide variety of lesion growth and arrest scenarios.
\end{abstract}

\maketitle
\section{Introduction}
MS affects about one million people worldwide and causes physical and cognitive disability. There are three types of MS: relapsing-remitting, secondary progressive and primary progressive, that differ in the dynamical patterns of disease progression. There are as yet no known cures for MS.  Patients with relapsing MS are currently treated with drugs  that exert immunomodulatory effects and slow the progression of the disease; there are no effective treatment options for the progressive forms of MS \cite{MS}.

MS is postulated to be a cell-mediated autoimmune disease directed against myelin components of the CNS. Myelin is an electrically insulating phospholipid layer that surrounds the axons of many neurons. The disease is characterized by both inflammatory immune responses and neuro-degeneration. The prevailing hypothesis on MS pathogenesis is that  auto-reactive T-lymphocytes, a cell type in  the immune system, orchestrate a complex cascade of events that cause blood brain barrier disruption and invasion of immunologically aggressive cells into the CNS.  However, the exact causes of MS still remain unknown \cite{MS1}. The long-term goals of this research are to develop disease models that can be used to evaluate therapeutic strategies for this disease and, in this report, the specific focus is on evaluating a network model for MS lesion dynamics. Literature survey indicates that network approaches have not been studied extensively for disease modeling in MS. 

\subsection{Previous work}
Conventional models for auto-immunity are premised on the  occurrence of defects in the immune system that cause it to turn against the host tissue. A defect-free immune system, in this world view,  purportedly only attacks pathogens, the external agents that cause illness or disease \cite{burnet}. However, an alternative viewpoint has  been advocated where auto-immunity is seen as the usual immune response, but  directed against those components of the body which, in normal conditions, are inaccessible to the immune system \cite{nossal,smir,matzinger,schwartz,nevo,schwartz1}. For example, in the {\em danger model}, developed by Matzinger  and collaborators \cite{matzinger}, it is posited that stressed and injured tissues can mediate immune responses through the generation of appropriate ``danger'' signals. This is as opposed to the activation through recognition of external pathogenic cell types from host tissue in the conventional models.  The concept of {\em comprehensive immunity}, developed by Nevo et al. \cite{schwartz, nevo} complements this alternate perspective; experimental results supporting their idea have also been reported \cite{schwartz1}. The present network model is inspired by the alternative viewpoint.
 
The key elements of the model consist of a pathological process that causes cellular damage and programmed cell death initiated through an inter-cellular signaling component. The programmed cell death deprives the pathological process of healthy tissue which is necessary for its propagation in space and time.  In this, it resembles the action of firemen who burn peripheral vegetation to contain forest fires. Inter-cellular signaling is a key feature of the model that allow pathologically damaged cells to propagate alarm signals and initiate programmed cell death.

\section{Model}

An undirected,  fixed radius random graph $G(n, r)$, with $n$ nodes (vertices) and radius of connectivity, $r$, is constructed to represent the CNS in this 2D network model. Fixed radius implies that nodes are connected only if they are within a distance of $r$. Biologically, the nodes of the graph can be viewed as representing cell bodies or functional units and the edges (bonds) of the graph can be viewed as axons or the interconnections between functional units.  

Let $d_i$ be the degree of the $i$th node, i.e. the number of edges attached to it. The health status of each edge, at time $t$, is indicated by its ``weight'', $w(j, t)$ ($j = 1, \ldots, d_i$), an integer number ranging from $0 \ldots w_{max}$. Edges with weight $w_{max}$ are fully functional or healthy units (as at the beginning of simulation), and those with weight zero, are  dead. 

In the pathological process, the edges are  damaged by lowering their weight by a single unit.  However, in the programmed cell death process, edge weights are directly reduced to zero. In the regeneration process,  edge weights are raised by a unit. 

The pathological and regeneration processes are driven by probabilistic  events wherein each edge in the affected region, in each time unit, has a certain probability \v{p}$_i^d$ (\v{p}$_i^r$) of getting damaged (regenerated). In the general case,  \v{p}$_i^d$ (\v{p}$_i^r$) is a column vector of length $w_{max}$ containing the transition probabilities from one state of health to another. Probability of programmed cell death,  $p_p$, is independent of the health status of the edge. 

The functional or health status of the $i$th node is the sum over its edge weights, $s_i (t) = \sum_{j=1}^{d_i} w(j, t)$. The maximum possible value of $s_i$ is denoted by $S_i$, which is realized when each $w(j, t) = w_{max}$. 

A node damaged by the pathological process generates an alarm signal when the ratio of its health status to the fully healthy state ($s_i(t)/S_i$)  falls below a threshold, \taual.  The signals received  at the $i$th node are summed  and  propagated further when the summed signal strength reaches $S_i$. The amplitude of the signal propagated along $j$th edge is equal to $w(j, t)$.

Programmed cell death is initiated at all the nodes where the propagated signals reach a  threshold \taubf. The accumulated alarm signals in the region of \prcell, a circular region around the activated node of radius proportional to a parameter \rb,  get reset to zero.   No additional signals are generated at these nodes, akin or otherwise, to the alarm signals generated in the pathologic process. 

The spread of the pathologic process is driven by the success rate in causing cellular damage. The fraction of edges damaged in a particular time step, among the total number of healthy edges visited, is computed  as the rate of damage due to the pathologic process, $R_I (t)$.  The radius of the region affected by the pathologic process increased or decreased according to the formula, $\alpha \times R_I (t) \times ROI_{t=0}$, where $ROI_{t=0}$ is the radius of the region at the center where the initial lesion is seeded. In a similar fashion, the region of programmed cell death was computed as \rb\ $\times ROI_{t=0}$. Final results are, thus, invariant with respect to the initial lesion size.

\section{Simulation}

In the simulations reported here, a two-state model with  $w_{max} = 1$ has been employed, i.e., there are no intermediate states of health and  the edges are either alive or dead. Additionally, the regeneration probability, \v{p}$^r_i$, was set to zero in order to focus exclusively on the effects of the interplay between the pathological and programmed cell death processes on lesion structure and dynamics. A few preliminary results using such a configuration was reported earlier \cite{km1}.

We have set $n = 400$ and chosen a uniform random distribution of points in the unit square  $[0, 1] \times [0,1]$ . The radius of connectivity was set to $r = 0.2$. All the results were also confirmed on a network of $n = 4000$, with $r = 0.06$. Average degree strengths of  the order of 10 are obtained in these configurations; degree distribution is gaussian. The pathological process was initiated at $t = 0$ in a region with $ROI_{t=0} = 0.05$ around the center at (0.5, 0.5); for $n = 4000$, $ROI_{t=0} = 0.015$.  

\begin{figure}[h]
\begin{center}
\includegraphics[width=3.3in]{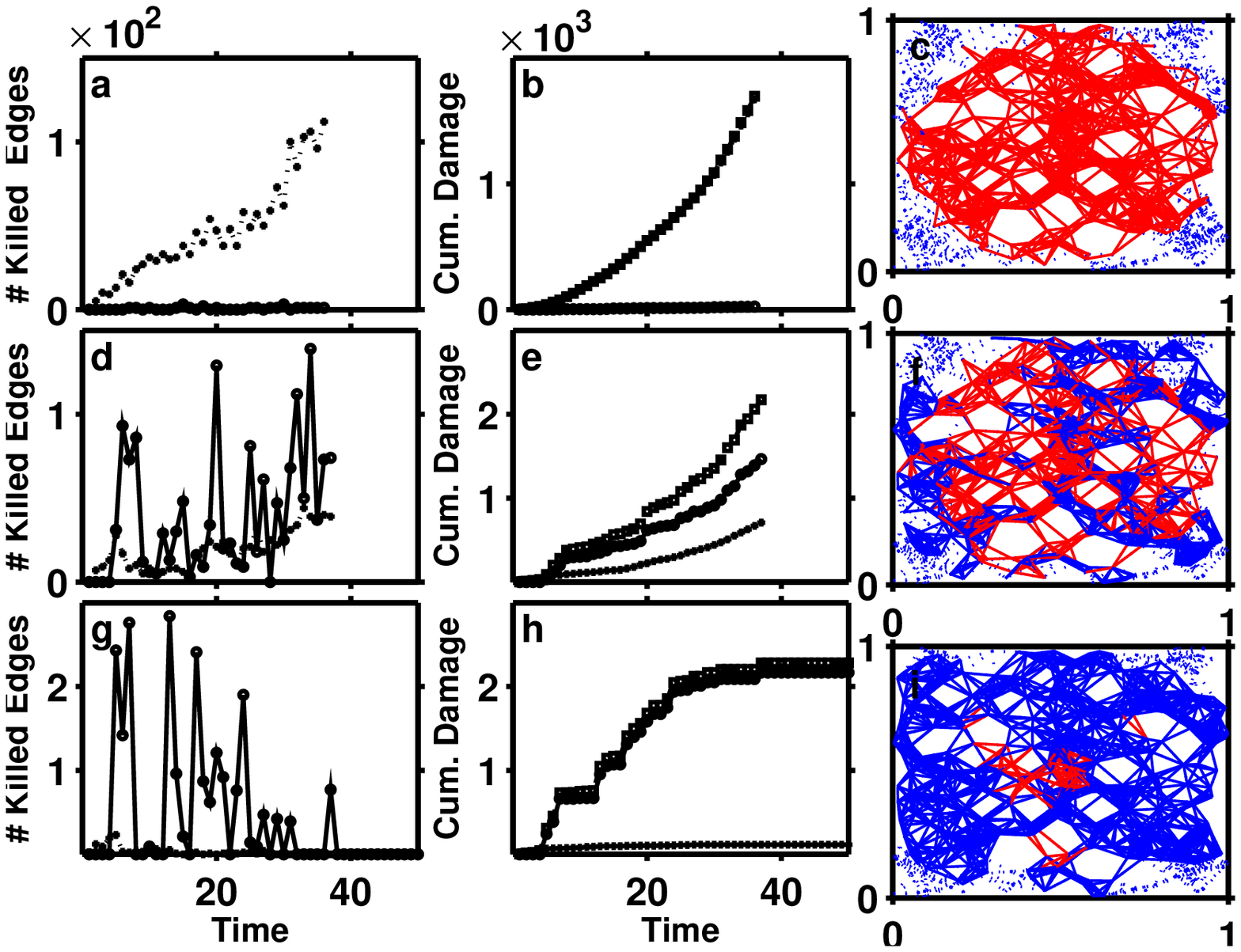}
\caption{Time course of damages to the system by the pathological and \prcell\ processes. The first column of panels show, separately, the instantaneous damages due to both the processes; dotted lines with asterisks indicate the damages due to the \patho, and bold lines with filled circles indicate damages due to \prcell. The second column of panels show the cumulative damages with time. Again, the damages effected through both the processes have been separately shown (same symbols as earlier), as also the sum total damages to the system (square symbol). The last column of panels (color online) show the state of the network at the end of the simulations; the dotted lines indicate healthy edges (axons), full (blue) lines indicate edges damaged due to \prcell, and dark (red) lines indicate edges damaged due to the \patho. For all the panels, \taubf\ $= 0.5$  and \taual\ $= 0.7$, while the \rb\ values, for each row, top to bottom, are 0.2, 0.8 and 1.5 respectively.}
\label{rinf1}
\end{center}
\end{figure}
A uniform probability  of pathologic damage  $p_d = 0.33$ was used, with $\alpha = 0.12$. We varied \taual, \taubf\ and \rb\ to identify the conditions under which the pathological process could be controlled by the programmed cell death. Larger values of \taubf\ indicate reduced sensitivity to the alarm signals, whereas a larger value of \rb\ indicates that a larger area near the alerted node is subjected to programmed cell death. In the case of \taual, larger values indicate quicker firing of alarm signals.

\section{Results}

Fig.~\ref{rinf1} shows the time series of damages caused to the system by both the \patho\ and the \prcell\ process. The first column of panels in the figure show the time course of instantaneous damages to the system. The middle coumn of panels show the time course of the cumulative damages to the system. The last column of panels show the final state of the network at the end of the simulations. 

There are three typical scenarios which are illustrated in Fig.~\ref{rinf1}, in the three rows from top to bottom.  Figs.~\ref{rinf1}a-c show a scenario where the \prcell\ is not of sufficient strength to significantly affect the \patho. Note that the instantaneous damages from \prcell\ are hardly ever above zero. Also, it  is seen from Fig.~\ref{rinf1}b that the contribution of \prcell\ to the sum total of damages is insignificant. This situation occurs with a suitable combination of low \taual, high \taubf\ and low \rb\ values. Figs.~\ref{rinf1}d-f shows a slightly more comples situation. In this case, \prcell\  is clearly the dominant effect. The instantaneous damages caused by both the processes are consistently nonzero (Fig.~\ref{rinf1}d) and the cumulative damages (Fig.~\ref{rinf1}e) continue to grow. Nevertheless, we see that the damage continues to spread.  In Figs.~\ref{rinf1}g-i, the \patho\ has been well controlled. The instantaneous damages have fallen to zero in Fig.~\ref{rinf1}g and the cumulative damages have levelled off even though the final state of the network shows that a large area of the network has been damaged.

\begin{figure}[h]
\begin{center}
\includegraphics[width=3.3in]{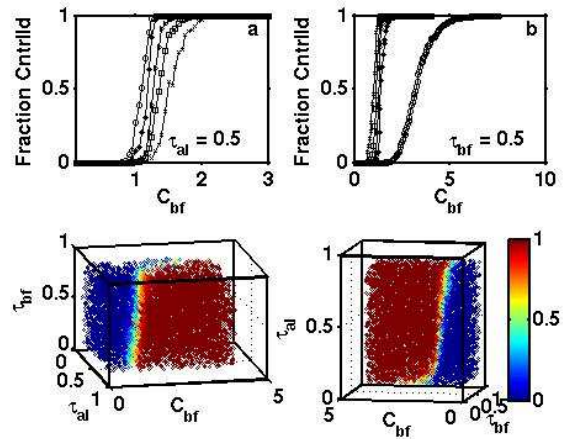}
\caption{Phase portraits of the dynamics. The $y$-axes in Figs.~\ref{rinf2}a-b and the gray (color online) scale in Figs.~\ref{rinf2}c-d indicate the fraction of the total number of simulations where the \patho\ was successfully arrested. In Fig.~\ref{rinf2}a, \taual\ has been held fixed at the shown value while the different curves are, from left to right, for \taubf\ $=$ 0.1, 0.3, 0.5, 0.7 and 0.9. In Fig.~\ref{rinf2}b, the different curves, from right to left, are for \taual\ $=$ 0.9, 0.7, 0.5, 0.3 and 0.1; \taubf\ $= 0.5$ is held fixed. The $S$-curves that result with increasing \rb\ values indicate the smooth nature of the transition in dynamical behavior in the parameter space.}
\label{rinf2}
\end{center}
\end{figure}

As seen from Fig.~\ref{rinf1}, the time series is stochastic. There are essentially two sources of randomness in the model. Firstly, the \patho\ is simulated by a binomial process wherein each edge visit will lead to successful damage if the generated random number falls below the value in \v{p}$_i^d$ for that edge. Secondly, the random network itself is generated by the random distribution of the $n$ points in the plane.  The complete picture of the transition from uncontrolled growth of the \patho\ to the situation where the \patho\ has been well arrested is seen in the parameter phase space graphs shown in Fig.~\ref{rinf2}, where an averaging has been effected over the two sources of randomness. The phase space diagrams are the results of averaging over ten different networks, with the dynamics averaged over a thousand iterations.

From Fig.~\ref{rinf2}, we see that the transition from uncontrolled \patho\ to arrested \patho\ is smooth as \rb\ is varied from low to high values. In Fig.~\ref{rinf2}a, \taual\ has been held fixed and the different curves, from left to right, are for different \taubf\ values, from 0.1 to 0.9 in steps of 0.2. In Fig.~\ref{rinf2}b, \taubf\ has been held fixed and the different curves are, from right to left, for \taual\ values ranging from 0.1 to 0.9, in steps of 0.2. We shall denote by \rbcr(\taual, \taubf) the critical value of \rb\ at which these $S$-curves attain a value of 1, i.e. all instances of simulations result in the growth of \patho\ being arrested. The combined picture in the three parameter space is presented in Figs.~\ref{rinf2}c-d, from two different perspectives. The three different scenarios presented, from top to bottom, in Fig.~\ref{rinf1} indicate, respectively, the three different parts, from left to right, of a typical $S$-curve of Figs.~\ref{rinf2}a-b.

\begin{figure}[h]
\begin{center}
\includegraphics[width=3.3in]{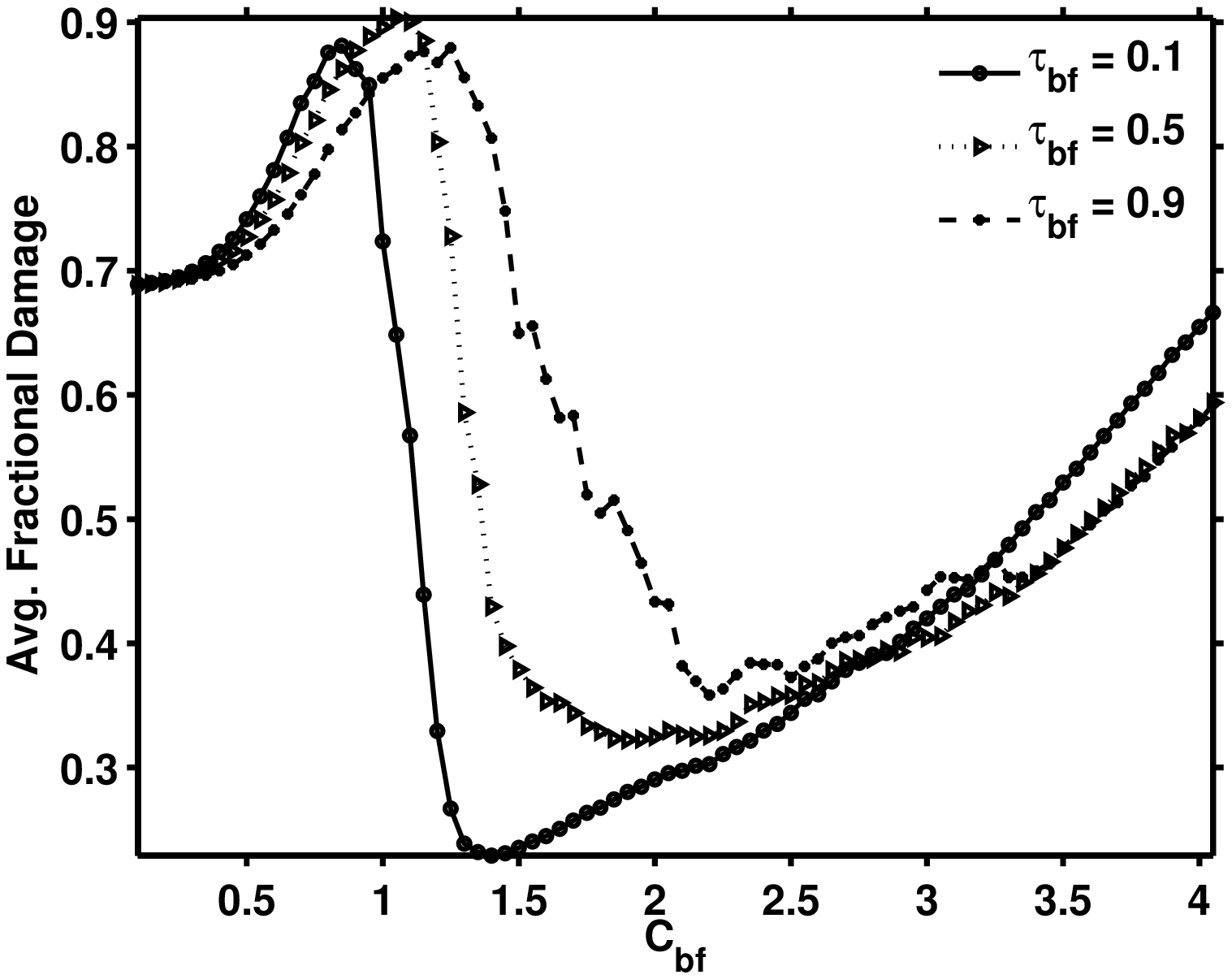}
\caption{Optimality in sum total damage to the system while effecting arrest of \patho\ is shown here. With increasing \rb\ values the minimum in sum total damage to the system occurs at about the same value as the critical value at which the fraction of simulations in which the \patho\ is arrested attains unity (cf. Fig.~\ref{rinf2}a). This minimum occurs at higher values of \rb\ with larger \taubf, and  the minimum value also shifts upward. A similar situation occurs with different \taual\ values (not shown here). Note that the $y$-axis values have been taken at $t = 20$.}
\label{rinf3}
\end{center}
\end{figure}
As seen from Fig.~\ref{rinf2}, \patho\ is always controlled if \rb\ $>$ \rbcr. Nevertheless, the sum total damage to the system is not the same for all values of \rb\ $>$ \rbcr; in fact, the damage is greater, the larger the value of \rb. Clearly, it is desirable to effect control of the \patho\ with the least sum total damage to the system. With this in mind, average fractional damages at different \rb\ values have been plotted in Fig.~\ref{rinf3}. Three different curves for three different \taubf\ values are shown in this figure; similar graphs can be constructed for different \taual\ values as well (not shown here; see \cite{km2}). These averages have been taken at $t = 20$ in each case. For \rb\ $<$ \rbcr\  values, the damages due to the \patho\ as well as sum total damage are still growing and have not become stationary at $t = 20$; for \rb\ $>$ \rbcr\  values, the averages have become stationary. Nevertheless, these curves indicate that the least sum total damage to the system, with \patho\ arrested, is obtained at \rb\ $=$ \rbcr.  For \rb\ $>$ \rbcr, the \prcell\ is clearly effecting more damage than is necessary to arrest the \patho. Since \rbcr\  depends on \taubf\ and \taual, it is not surprising to see that (cf. Fig.~\ref{rinf2}a and Fig.~\ref{rinf3}) lesser damage results when \taubf\ is small. 

It has already been seen (cf. Figs.~\ref{rinf1}d-e) that arrest of the \patho\ doe not necessarily occur if the damage due to the \prcell\ process is greater than the \patho. What is necessary \cite{km2} is that the \prcell\ process be able to encircle the region affected by the \patho, and, furthermore, be able to create an envelope region of sufficient thickness to offset its likely growth factor, $\alpha \times R_I(t) \times ROI_{t=0}$. This is achieved in all instances of simulation when \rb\ $>$ \rbcr. Currently, mathematical analysis of this feature is being carried out to establish the relationship of \rbcr\ with \taual\ and \taubf, and the results will be reported soon.

\section{Conclusions}

A physically motivated 2D network model was developed for the CNS and employed to study  the process of lesion formation and spread in  MS. Inter-cellular signalling of distress by the damaged cells is a key feature of the model which leads to \prcell\ getting activated in an attempt to arrest the lesion progress. The model demonstrates that the spread of the pathologic process can be  arrested by programmed cell death when the geometry of the damage inflicted by the latter leads to an envelope, of sufficient thickness, being created encircling the area of \patho. Such an envelope of dead cells deprives the \patho\ of healthy cells which can sustain its growth. The model shows a smooth transition, as parameters are varied, from the situations of run-away \patho, through aggravated damage to the system caused by unsuccessful firing of \prcell, to the creation of successful envelope around the \patho. 

The model complements the alternate viewpoint on auto-immunity which posits that cells and tissues signal distress and activate the immune system. Such a viewpoint circumvents the need for the immune system to store information about likely pathogens and, also, makes it capable of acting in instances of cellular damage resulting from non-pathogenic causes. Further study of the model along with identification of the possible biological constituents should enable comparisons with experiments and a more detailed exposition.

\section*{Acknowledgements}
This work was carried out during a period spent at the Department of Pharmaceutical Sciences, State University of New York at Buffalo (UB), USA. Encouragement, discussions and funding from Dr. Murali Ramanathan is acknowledged. Use of the compuational facilities at the Center for Computational Research, UB facilitated the work.

\end{document}